# Prediction of New Ground State Crystal Structure of $Ta_2O_5$

Yong Yang[1*] and Yoshiyuki Kawazoe[2,3]

1. Key Laboratory of Materials Physics, Institute of Solid State Physics, Chinese Academy of Sciences, Hefei 230031, China.
2. New Industry Creation Hatchery Center (NICHe), Tohoku University, 6-6-4 Aoba, Aramaki, Aoba-ku, Sendai, Miyagi 980-8579, Japan.
3. Department of Physics and Nanotechnology, SRM University, Kattankulathurm, 603203, TN, India.

Tantalum pentoxide ($Ta_2O_5$) is a wide-gap semiconductor which has important technological applications. Despite the enormous efforts from both experimental and theoretical studies, the *ground state crystal structure* of $Ta_2O_5$ is not yet uniquely determined. Based on first-principles calculations in combination with evolutionary algorithm, we identify a triclinic phase of $Ta_2O_5$, which is energetically much more stable than any phases or structural models reported previously. Characterization of the static and dynamical properties of the new phase reveals the common features shared with previous metastable phases of $Ta_2O_5$. In particular, we show that the *d*-spacing of ~ 3.8 Å found in the X-ray diffraction (XRD) patterns of many previous experimental works, is actually the radius of the second Ta-Ta coordination shell as defined by radial distribution functions.

*Corresponding Author (wateratnanoscale@hotmail.com; yyang@theory.issp.ac.cn)

**I. INTRODUCTION**

As a versatile wide-gap semiconducting material, tantalum pentoxide ($Ta_2O_5$) has attracted a lot of interests of research in the past decades. The intrinsic excellent dielectric properties make $Ta_2O_5$ a good candidate material to replace $SiO_2$ as the insulating layer for advanced electronic devices such as non-volatile memories [1-5] and organic transistors [6, 7]. Owing to its high refractive index, $Ta_2O_5$ has applications in the coating material of optical devices [8, 9] and the instruments for ultraprecise measurement such as the Laser Interferometer Gravitational-wave Observatory (LIGO) [10-12]. In the detection and observation of gravitational waves made by LIGO [13], alternating thin layers of $SiO_2$ and $Ta_2O_5$ (doped with $TiO_2$) serve as mirror coatings for the test masses of the detectors [10-12]. Additionally, as a transition metal oxide, $Ta_2O_5$ also finds its place in other applications such as corrosion resistant coatings [14, 15] and catalyst for electro- and photocatalysis [16-19].

Previous experimental studies have established that, $Ta_2O_5$ undergoes a phase transition at $T \sim 1593$ K (1320 °C), from the low-temperature phase (L-$Ta_2O_5$) to the high-temperature phase (H-$Ta_2O_5$) [20]. In the research works that followed, many efforts are devoted to studying the crystallographic structures of both L-$Ta_2O_5$ and H-$Ta_2O_5$, which still remain an issue of debate [21, 22]. Our work will focus on the atomic structures of the low-temperature phase of $Ta_2O_5$, L-$Ta_2O_5$, which is more relevant to technological applications. Due to the difficulties of growing high-quality single crystals of $Ta_2O_5$, the structural information provided by powder X-ray diffraction (XRD) is very limited. Indeed, the crystal structures are found to be critically dependent on the conditions (e.g., temperatures, pressures) of synthesis and the method of analysis [23]. A number of phases and/or structural models have been suggested for L-$Ta_2O_5$. Based on the data of powder XRD, Stephenson and Roth proposed the 11-formula-units ($Z = 11$, 22 Ta and 55 O atoms) orthorhombic model (referred to as $L_{SR}$) [24]. Later, the Hummel *et al.* proposed the *T*-phase (orthorhombic, space group: *Pmm*2) which is however nonstoichiometric in chemical components (unit cell contains 24 Ta and 62 O atoms) [21]. Grey, Mumme, and Roth synthesized a new phase of L-Ta$_2$O$_5$ (referred to as L$_{GMR}$) whose unit cell is monoclinic with Z = 19 [25]. For the β-phase of L-Ta$_2$O$_5$, at least two different models with orthorhombic unit cells are found: the one (space group: *Pccm*) proposed by Aleshina and Loginova using Rietveld analysis of XRD pattern (referred to as β$_{AL}$) [26], and the one proposed by Ramprasad (referred to as β$_R$) [27], which is actually a simplified version of the L$_{SR}$ model. Another orthorhombic phase, namely, the λ-model has been recently proposed with the space group *Pbam* [28]. The hexagonal unit cells of Ta$_2$O$_5$ are also reported by experimental works [21, 29]; however, the unit cell parameters and internal atomic coordinates of the phase are not fully determined until the work by Fukumoto and Miwa, which they called δ-Ta$_2$O$_5$ [30]. Two high-pressure phases, B-Ta$_2$O$_5$ (also known as $\varepsilon$-Ta$_2$O$_5$ [31]) and Z-Ta$_2$O$_5$ are synthesized at *P* = 8 GPa and *T* = 1470 K [32]. The B-Ta$_2$O$_5$ is found to be stable at atmospheric pressure and low temperatures. From the formation energies given by a recent work based on density functional theory (DFT) calculations [22], the stability of different phases of Ta$_2$O$_5$ is ranked in the order of B > λ > L$_{SR}$ > β$_R$ > L$_{GMR}$ > δ > β$_{AL}$. On the other hand, phonon calculations based on DFT show that the two β-phases (β$_{AL}$, β$_R$), and the δ-phase are dynamically unstable [22, 33].

In spite of these efforts, the *ground state crystal structure* of L-Ta$_2$O$_5$ is still elusive. Recent theoretical works based on the L$_{SR}$ or the λ-phase have shown that, the ground state atomic structures of L-Ta$_2$O$_5$ can be energetically highly degenerated, and the creation of dilute oxygen vacancies can induce long-ranged perturbations on the atomic positions [34-36]. The concept of infinitely adaptive crystal structure [37] may account for such uncertainties or varieties of crystal structures, and leaves open the possibility of finding new phases/polymorphs of Ta$_2$O$_5$. In this work, we predict, based on *ab initio* evolutionary structure searches, a possible triclinic phase of Ta$_2$O$_5$, which is found to be more stable than any phases of Ta$_2$O$_5$ reported in literatures.

## II. COMPUTATIONAL METHODS

Structure searches are performed using the *ab initio* evolutionary algorithm implemented in the USPEX package, which has demonstrated its reliability in identifying new phases of bulk materials [38, 39]. The evolutionary simulations are carried out in the $Ta_2O_5$ systems whose unit cells contain two formula units (Z = 2). The first generation is produced randomly (119 structures), and the following generations are produced by heredity and lattice mutation (each generation contains 40 ~ 50 structures). Good convergence in the total energies is obtained after 19 generations (Fig. S1), surveying totally 1035 different crystal structures. During the structure search, all the local structural optimizations are done by the VASP code [40, 41], using a plane wave basis set and the projector-augmented-wave (PAW) potentials [42, 43], and uniform k-meshes with a spacing of ~ $0.06 \times 2\pi Å^{-1}$ for integration. The exchange-correlation interactions of electrons are described by the PBE type functional [44]. The energy cutoff for plane waves is 600 eV. The total energy of each configuration converges to a level of less than 1 meV/unit cell. After the evolutionary simulations, we analyzed the results by selecting a few of the lowest-energy structures (Generation 11 to 19) and optimized the unit cell and atomic structures again using VASP with a denser $10 \times 10 \times 4$ k-mesh, which ensures the total energy to converge to a level of less than 0.1 meV/unit cell. The set of the lowest-energy structures under consideration are well converged after re-optimization. The obtained new phase of $Ta_2O_5$ has a triclinic unit cell with the dimensional parameters (lengths and angles) as: $a$ = 3.89 Å, $b$ = 3.89 Å, $c$ = 13.38 Å, $α$ = 81.77°, $β$ = 98.25°, $γ$ = 89.67°, and space group $P$1 (atomic coordinates are provided in the Table SI). We have also optimized the structures of B, λ, $L_{SR}$, δ and $β_{AL}$ phase of L-$Ta_2O_5$, and calculated their total energies using VASP to make a comparison with the new triclinic phase, which we call the γ phase hereafter for simplicity. The k-meshes for the calculations of B, λ, $L_{SR}$, δ and $β_{AL}$ phase are $2 \times 4 \times 4$, $4 \times 4 \times 8$, $4 \times 2 \times 4$, $6 \times 6 \times 12$, $6 \times 12 \times 6$, respectively. All the k-meshes are generated by using the Monkhorst-Pack scheme [45]. The phonon dispersion and vibrational density of states (VDOS) of the γ phase are computed using the Quantum Espresso program [46], with the computational details provided in the Supplemental Material.

## III. RESULTS AND DISCUSSION

## A. Comparison with the previously identified phases of L-Ta$_2$O$_5$

The crystal structures of six phases of Ta$_2$O$_5$: $\beta_{AL}$, $\delta$, L$_{SR}$, B, $\lambda$ and $\gamma$ are schematically shown in Fig. 1. Basically, the crystals of Ta$_2$O$_5$ are built up by a group of polyhedrons centered at Ta atoms with the vertexes of polyhedrons being O atoms [22, 27, 34]. From the number of O atoms that bonded with the central Ta, the polyhedrons can be classified into three types: octahedron (TaO$_6$), pentagonal bipyramid (TaO$_7$), and hexagonal bipyramid (TaO$_8$). The $\beta_{AL}$, B, $\lambda$ and $\gamma$ phase are solely formed by distorted octahedrons. The hexagonal $\delta$ phase is formed by edge-sharing octahedrons and hexagonal bipyramids, while the L$_{SR}$ phase is constructed by edge and corner-sharing octahedrons and pentagonal bipyramids (Fig. 1). The optimized lattice parameters of the six Ta$_2$O$_5$ phases, the mass densities and their ground state energies with respect to the $\gamma$ phase are tabulated in Table I. As expected, the PBE functional tends to overestimate the value of lattice parameters (by ~ 4.8% at most) while the PBEsol type functional [47] better describes the equilibrium geometries (error by ~ 3.5% at most, see Table SII). The PBE functional is chosen here for most calculations because it performs better than PBEsol in the calculation of energetic parameters [47, 48], which is our first concern for ranking the stability of different Ta$_2$O$_5$ phases. For the known Ta$_2$O$_5$ phases ($\beta_{AL}$, $\delta$, L$_{SR}$, $\lambda$, B), the order of stability is found to be the same as previous studies [22]. To make a comparison, the relative energies of the optimized structures of the six phases calculated using the PBE [44], PBEsol [47], PW91 [49], HSE06 [50-52] and PBE0 [53] type functional are presented in Table I. It is clear that the $\gamma$-Ta$_2$O$_5$ stands as the most stable one for calculations using all the five types of functionals. Additionally, the first, second and third energetically favored phases, $\gamma$, B, and $\lambda$-Ta$_2$O$_5$, are built up only by octahedrons (TaO$_6$). This is in line with a recent qualitative analysis on B and $\lambda$-Ta$_2$O$_5$ [22]. For $\gamma$-Ta$_2$O$_5$, the band gap calculated by PBE functional is ~ 2.26 eV, and is increased to ~ 3.75 eV and 4.51 eV (Fig. S2), for calculations using the HSE06 [50-52] and PEB0 [53] type hybrid functionals, respectively. The band gap predicted by HSE06 is close to the value of $\lambda$-Ta$_2$O$_5$ (3.7 eV) [22, 28] but smaller than that of B-Ta$_2$O$_5$ (4.7 eV) [22]. The HSE06 band gaps of both $\lambda$ and $\gamma$-Ta$_2$O$_5$ are comparable with the experimental value, which is ~ 4 eV [54].

### B. Possible pressure-induced phase transformation

The notable difference between the mass densities (Table I) of B, λ and γ-Ta$_2$O$_5$ implies that phase transformation may be induced by applied pressures. We firstly calculated the total energies $E$ for a set of volumes $V$, and then deduced the analytic function $E(V)$ by least-squares fitting the $E$-$V$ data to Murnaghan's equation of state (EOS) [55, 56]: $E(V) = \frac{B_0 V}{B_0'}\left[\frac{(V_0/V)^{B_0'}}{B_0'-1} + 1\right] + E(V_0) - \frac{B_0 V_0}{B_0'-1}$, where $B_0$ and $B_0'$ are respectively the bulk modulus and its pressure derivative at equilibrium volume $V_0$ at which $E(V)$ reaches its minimum. The parameters $B_0$ and $B_0'$ for the EOS of the three phases are listed in Table SIII. At $T = 0$ K, phase transformation occurs when the enthalpy $H = E + PV$ of two phases equals. It can be shown that pressure-induced phase transformation takes place along the common tangent line connecting the $E(V)$ curves of the two phases, with the transition pressure ($P_t$) given by negative value of the slope of the tangent line, and the corresponding transition volume ($V_t$) given by the point of tangency. The $E(V)$ curves and the tangent lines related to the phase transformation from γ-Ta$_2$O$_5$ to B and λ-Ta$_2$O$_5$ are shown in Fig. 2, and the transition pressures and volumes are listed in Table II. The transition pressure is ~ 1.78 GPa for the transformation from γ-Ta$_2$O$_5$ to B-Ta$_2$O$_5$, and is ~ 4.16 GPa for the transformation from γ-Ta$_2$O$_5$ to λ-Ta$_2$O$_5$. Experimentally, B-Ta$_2$O$_5$ was synthesized in a high-pressure chamber at $P = 8$ GPa and $T = 1470$ K [32]. Therefore, it is possible that in the initial stage of compression at lower pressure and temperature conditions, γ-Ta$_2$O$_5$ exists as a precursor phase prior transforming into the B phase. In the following studies, detailed comparison on the structural properties of these three phases will be presented.

### C. Analysis of XRD patterns and radial distribution functions (RDF)

Figure 3(a) shows the simulated X-ray diffraction (XRD) pattern of γ-Ta$_2$O$_5$. The strongest reflection appears at the (010), (100), (011), and (10$\bar{1}$) diffraction planes with the incident angle $2\theta$ varying from ~ 23.11° to 23.12°, and the $d$-spacing from ~ 3.84 Å to 3.85 Å. The smallest incident angle that produces a clear diffraction locates at $2\theta = 13.51°$, with the index (002) and a $d$-spacing of ~ 6.55 Å, which is accordingly the largest separation between each pair of diffraction planes. The second and third strongest diffraction peaks are found at $2\theta = 33.26°, 27.21°$; with the $d$-spacing ~ 2.70 Å, 3.27 Å, and the corresponding indexes (103), (004), respectively. As shown below, the strongest diffraction peak is associated with the second Ta-Ta coordination shell as defined by radial distribution function.

To study the dynamic arrangement of the neighboring atoms around Ta, the center of the TaO$_6$ octahedrons, we have performed *ab initio* molecular dynamics (MD) simulations at 300 K (detailed in Supplemental Texts) for γ, B, and λ-Ta$_2$O$_5$, and calculated the radial distribution function (RDF) by averaging over all the atomic configurations involved within the simulation time. The calculated RDFs $g_{TaO}$ and $g_{TaTa}$ of γ-Ta$_2$O$_5$ are displayed in Figs. 3(b) and 3(c). The first and second RDF peaks of $g_{TaO}$ ($g_{TaTa}$) correspond to the first and second Ta-O (Ta-Ta) atomic coordination shells, respectively. Broadening of the first peak of $g_{TaO}$ corresponds the variation range of Ta-O bond lengths, whose averaged value (~ 2.00 Å) is approximately at the peak position ($R_{1P}$), as indicated in Fig. 3(b). The geometric parameters associated with the RDFs of γ-Ta$_2$O$_5$ are listed in Table III, together with parameters related to the RDFs of B (Fig. S3) and λ-Ta$_2$O$_5$ (Fig. S4). Comparing the $g_{TaO}$ of B and λ-Ta$_2$O$_5$ with γ-Ta$_2$O$_5$, one sees that the averaged lengths of Ta-O bonds are all located at ~ 2 Å, in good agreement with available experimental data for TaO$_6$ octahedrons [24, 32].

For $g_{TaTa}$ which describes the spatial distribution of TaO$_6$ octahedrons, the first and second peak of γ-Ta$_2$O$_5$ locates at $R_{1P}$ ~ 3.32 Å and $R_{2P}$ ~ 3.84 Å, respectively. As found in Table III, the values of $R_{1P}$ and $R_{2P}$ of the three phases are close to each other: where the variation of $R_{1P}$ and $R_{2P}$ can be described by $R_{1P}$ ~ 3.30 ± 0.02 Å and $R_{2P}$ ~ 3.80 ± 0.08 Å. The value of $R_{2P}$ ~ 3.80 ± 0.08 Å, is measured as one of the lattice parameters ($c$-axis) in a number of experimental studies on different Ta$_2$O$_5$ phases [21, 23-25, 29, 57, 58]. In the case of γ-Ta$_2$O$_5$, it corresponds to the length of $a$-axis and $b$-axis. Therefore, the physical meaning of such a *characteristic length* is evident: It is the radius of the second Ta-Ta coordination shells. Considering the fact that Ta atoms contain much more electrons than O atoms and consequently contribute dominantly to the XRD signals, it is natural to understand that it is the periodic arrangement of Ta atoms that determines the lattice parameters of $Ta_2O_5$. Indeed, this is supported by the XRD pattern shown in Fig. 3(a), in which the strongest diffraction is induced by a bundle of planes with the *d*-spacing of ~ 3.84 Å to 3.85 Å. Moreover, strong diffraction peaks at the corresponding incident angle $2\theta$ ~ 23° under the Cu *K*α radiation (wave length $\lambda$ ~ 1.541 Å) were observed in previous experimental measurements in different $Ta_2O_5$ polymorphs [18, 23, 31, 32, 59, 60]. The analysis above reveals that similar local atomic coordination structures are shared by different $Ta_2O_5$ polymorphs.

### D. Comparison of the local bonding structures

The $TaO_6$ octahedrons constructing B, λ and γ-$Ta_2O_5$ are schematically shown in Figs. 4(a)-(c). Similarity of local Ta-O bonding structures is found in the Ta-O bond lengths, the surface areas and volumes of the octahedrons. The averaged Ta-O bond length is 2.02, 2.00, and 2.01 Å; the surface area is 27.78, 27.30, and 27.34 $Å^2$; and the volume is 16.02, 15.24, and 15.52 $Å^3$, for B, λ and γ-$Ta_2O_5$, respectively. For each phase, the difference between the geometric parameters of every two octahedrons is found to be less than 1% and therefore can be regarded as the same. On the other hand, one can find the difference between the coordination O atoms, which can be divided into two types: the ones bonded with two Ta (referred to as $O_{2c}$), and the ones bonded with three Ta (referred to as $O_{3c}$). From the stoichiometry that every Ta is associated with 2.5 O atoms on average, one has the equalities: $n \times 1/3 + m \times 1/2 = 2.5$; $n + m = 6$, where *n* is the number of $O_{3c}$ and *m* the number of $O_{2c}$. It follows that $n = m = 3$, which means that the number of the two types of O atoms are equal for the $TaO_6$ polyhedrons. On the other hand, the coordination structures of $TaO_6$ octahedrons of the three phases are distinct: Around the central Ta, there are six geometrically different O atoms in B-$Ta_2O_5$, while only five different O in λ-$Ta_2O_5$ and four different O in γ-$Ta_2O_5$. The six-coordination of λ and γ-$Ta_2O_5$ is achieved by periodic extension of the crystal unit cells: the self-donated coordination O, i.e., the O "V" marked in Fig. 4(b) and O "I" and "IV" in Fig. 4(c). It is worthwhile to note here that, although the number of $O_{2c}$ and $O_{3c}$ is equal for the $TaO_6$ octahedron, the total number of $O_{2c}$ and $O_{3c}$ is different in the unit cell, whose ratio is 3:2 for the three phases.

Figures 4(d)-(f) show the neighboring octahedrons around an indicated $TaO_6$ octahedron, within the first and second Ta-Ta coordination shells as defined by $g_{TaTa}$ [Fig. 3(c), Fig. S3(b), Fig. S4(b)]. The first and second coordination octahedrons are considered here due to the reasons: Firstly, the peak positions $R_{1P}$ and $R_{2P}$ only differ by ~ 0.5 Å or less; and secondly, the boundary between the two coordination shells are flexible for B and $\lambda$-$Ta_2O_5$, while only the sum of Ta-Ta coordination numbers (CNTa) of the two shells is invariant. The value of CNTa is 8 for both B and $\lambda$-$Ta_2O_5$ and is 7 for $\gamma$-$Ta_2O_5$. The difference of CNTa originates from the positions of $O_{3c}$ atoms as shown in Figs. 4(g)-(i). In principle, the maximum of CNTa that one $TaO_6$ can have is $n \times 2 + m \times 1 = 3 \times 2 + 3 \times 1 = 9$, by considering that one $O_{3c}$ can contribute two neighboring Ta and one $O_{2c}$ contribute one Ta at most. When two $O_{3c}$ locate nearby and share one neighboring Ta with each other, as indicated in Figs. 4(g)-(h), the value of CNTa is reduced by one, i.e., CNTa = 8; when the three $O_{3c}$ locate side-by-side as indicated in Fig. 4(i), CNTa is reduced by two, i.e., CNTa = 7. The smaller value of CNTa helps to understand the lower mass density of $\gamma$-$Ta_2O_5$.

**E. Analysis of Phonons**

We go on to study the dynamical properties of $\gamma$-$Ta_2O_5$ by calculating the phonon spectrum based on density functional perturbation theory (DFPT) [61]. Compared to the direct method in which a large supercell may be employed [62], the DFPT method can study the phonon properties within the primitive cell. Figures 5(a)-(b) show the phonon dispersion along the $\Gamma$-X and $\Gamma$-L lines, and the vibrational density of states (VDOS) of phonons is shown in Fig. 5(c). The dynamical stability of $\gamma$-$Ta_2O_5$ is therefore confirmed by the absence of imaginary frequency points in the vibrational spectrum. Furthermore, the thermal stability of $\gamma$-$Ta_2O_5$ is also demonstrated by the regular and nearly equal-amplitude fluctuations of total energies around a constant expected value as obtained by our MD simulations at 300 K (Fig. S5).

In previous works [59, 60, 63], the vibrational modes with the wavenumber $\tilde{v} > 800$ cm$^{-1}$ are assigned to the internal Ta-O stretching motions of O$_{2c}$ atoms, the modes at the region of 400 cm$^{-1}$ < $\tilde{v}$ < 800 cm$^{-1}$ are assigned to the stretching motions of O$_{3c}$ atoms, the modes at the region of 150 cm$^{-1}$ < $\tilde{v}$ < 400 cm$^{-1}$ are attributed to the deformation motions of O$_{2c}$ and O$_{3c}$ with respect to the bonded Ta atoms, and the low frequency part $\tilde{v}$ < 150 cm$^{-1}$ are attributed to the external motions of TaO$_n$ polyhedrons or Ta$_x$O$_y$ clusters. We examine here the applicability of these assignments by studying the characteristics of vibrational modes of γ-Ta$_2$O$_5$ at the Γ-point (long-wavelength limit, VDOS shown in Fig. S6), which contribute majorly to the signals of Raman or infrared spectroscopy. The mode with the highest vibrational energy at Γ-point locates at $\tilde{v}$ = 1007 cm$^{-1}$ (Fig. S6), very close to the VDOS peak $\tilde{v}$ ~ 1009 cm$^{-1}$ indicated in Fig. 5(c). The polarization vector of this mode is shown in Fig. 5(d), which describes the direction and magnitude of atomic vibrations. It is clear that this mode is mainly due to the stretching motions of O$_{2c}$ in the two neighboring TaO$_6$, vibrating nearly along the *c*-axis with opposite directions. This is in line with the assignment of Raman spectroscopy measurement on the L$_{SR}$ phase of Ta$_2$O$_5$ [60]. Correspondence with experiments is also found at the middle and low frequency part, for the vibrational modes at the Γ-point. The modes with $\tilde{v}$ ~ 510 cm$^{-1}$ and 570 cm$^{-1}$ (Fig. S6) have been reported by measurements using Fourier-transform infrared spectroscopy (FTIR) on the thin films of Ta$_2$O$_5$ [63]. From their polarization vectors [Figs. S7(a)-(b)], these two modes are mainly associated with the vibrations of O$_{3c}$ atoms. The mode with the wavenumber $\tilde{v}$ ~ 255 cm$^{-1}$ is recorded in the Raman spectra of different Ta$_2$O$_5$ phases [18, 59, 64]. In γ-Ta$_2$O$_5$, this mode is largely due to the vibrations of O$_{2c}$ atoms along the *c*-axis, which leads to the bending of Ta-O-Ta bonds [Fig. S7(c)]. The lowest mode originating mainly from the vibrations of O$_{2c}$ is located at the wavenumber $\tilde{v}$ ~ 91 cm$^{-1}$, which corresponds to the collective vibrations of O$_{2c}$ along the *b*-axis [Fig. S7(d)]. For the vibrational modes with $\tilde{v}$ < 90 cm$^{-1}$, the motions of Ta atoms start to come into play. For instance, the two modes at $\tilde{\nu}$ = 34 cm$^{-1}$, 35 cm$^{-1}$ were reported in previous Raman spectroscopy measurements [59], which correspond to the collective vibrations of Ta and O atoms of γ-Ta$_2$O$_5$ (O$_{2c}$ and O$_{3c}$) in the *ab* plane of the unit cell (Fig. S8). On the other hand, the vibrational spectrum of Ta$_2$O$_5$ can also be calculated by Fourier transform of the velocity autocorrelation function, which is readily deduced from the atomic trajectories recorded in MD simulations. Compared with the VDOS obtained by DFPT method, slight softening is observed in the Ta-O stretching modes of the vibrational spectrum (Fig. S9). This is mainly due to the different types of exchange-correlation functionals employed in the DFPT (Perdew-Zunger (abbr.: PZ) type, see Supplemental Texts) and MD simulations (PBE type), which lead to small difference in the optimized atomic structures and interatomic forces determined for vibrational frequency calculations. For instance, our DFPT calculations at Γ-point found that, the frequency of the highest and the second highest vibrational mode is ~ 1007 cm$^{-1}$ and 910 cm$^{-1}$ by PZ type functional, while it is ~ 967 cm$^{-1}$ and 839 cm$^{-1}$ by PBE type, respectively. Another reason may be the different temperatures considered for the system: 0 K in DFPT and 300 K in MD (mainly from Γ-point vibrations).

It should be noted here that the vibrational density of states is contributed from the phonons not only at Γ-point, but also at the other wave vectors (nonzero q-point) generated by the q-mesh employed in our DFPT calculations. Therefore, some of the vibrational modes at Γ-point appear as sides or shoulders of the peaks shown in Fig. 5(c). Nevertheless, comparison with the experimental data from different Ta$_2$O$_5$ polymorphs has demonstrated that, the common features shared by the local bonding structures (O$_{2c}$, O$_{3c}$) are reflected in some typical vibrational modes of the phonon spectra, despite the difference in mid- and long-range order which gives rises to different crystal structures. Indeed, previous measurements observed only minor difference between the Raman spectra of ceramic and crystalline Ta$_2$O$_5$ at room temperature [59]. Comparison with the RDFs calculated in amorphous Ta$_2$O$_5$ [65] also shows that, the upper limit of radii of the first and second Ta-O and Ta-Ta coordination shells in crystalline and amorphous Ta$_2$O$_5$ are approximately the same values.

Within the harmonic approximation, the vibrational contribution to the free energy can be written as [61] $F_{vib} = k_B T \sum_i \log\{2\sinh[\hbar\omega_i/(2k_B T)]\}$, where $T$ is temperature, $\omega_i$ is phonon frequency, $\hbar$ and $k_B$ have the usual meanings. Under cryogenic conditions, the vibrational contribution to free energy is negligible. However, at temperatures well above 300 K, the entropy term due to atomic vibrations increases with elevating temperatures and therefore plays an important role in the total free energy of the system. As a result, the order of stability of different phases of $Ta_2O_5$ may change and the temperature-induced phase transition will take place. This is the topic of future research.

## IV. CONCLUSIONS

In conclusion, using *ab initio* evolutionary algorithm, we report a possible new phase of $Ta_2O_5$ with triclinic unit cells. The triclinic phase is found to be more stable than any existing phases or structural models at low temperature. Detailed analysis and comparison with experimental data reveal that, common features present not only in the local bonding structures (e.g., the Ta-O bond lengths, the Ta and O local coordination structures, and the *d*-spacing ~ 3.8 Å in XRD pattern) of the triclinic phase and the other phases, but also in the dynamical properties such as the phonon spectra. Pressure-induced phase transformation is predicted to occur between the triclinic phase and the B and λ phase of $Ta_2O_5$ at pressures of several GPa. The results are expected to stimulate future experimental verifications and in-depth theoretical studies.

## ACKNOWLEDGMENTS

We are grateful to the staff of the Hefei Branch of Supercomputing Center of Chinese Academy of Sciences; and the crew of Center for Computational Materials Science of the Institute for Materials Research, Tohoku University for their support of the SR16000 supercomputing facilities. We would like to thank Professor Wenguang Zhu for reading and helpful comments on the manuscript. Yong Yang acknowledges support from the National Natural Science Foundation of China (Grant No. 11474285).

**TABLE I.** Calculated lattice and energetic parameters of the six $Ta_2O_5$ phases, where $α$, $β$, $γ$ are the angles between the cell edges $b$ and $c$, $a$ and $c$, and $a$ and $b$, respectively. Z is the number of $Ta_2O_5$ units in the unit cell, and $ρ$ is mass density. The available experimental data (with the subscript expt) of cell lengths are listed for comparison. $ΔE$ is ground state energy difference (per formula unit) relative to the $γ$ phase, calculated using the PBE, PBEsol, PW91, HSE06 and PBE0 type exchange-correlation functionals. Due to the extremely heavy computational burden, the k-meshes employed in the HSE06 and PBE0 type hybrid functional calculations are 8×8×2, 2×4×4, 4×4×8, 4×2×4, 4×4×8, and 4×6×4, for the $γ$, B, $λ$, $L_{SR}$, $δ$ and $β_{AL}$ phase, respectively.

| | | $β_{AL}$ | $δ$ | $L_{SR}$ | $λ$ | B | $γ$ |
|---|---|---|---|---|---|---|---|
| $a$ (Å) | | 6.52 | 7.33 | 6.33 | 6.25 | 12.93 | 3.89 |
| $a_{expt}$ (Å) | | [a]6.217 | -- | [b]6.198 | -- | [c]12.7853 | -- |
| $b$ (Å) | | 3.69 | 7.33 | 40.92 | 7.40 | 4.92 | 3.89 |
| $b_{expt}$ (Å) | | [a]3.677 | -- | [b]40.290 | -- | [c]4.8537 | -- |
| $c$ (Å) | | 7.78 | 3.89 | 3.85 | 3.82 | 5.59 | 13.38 |
| $c_{expt}$ (Å) | | [a]7.794 | -- | [b]3.888 | -- | [c]5.5276 | -- |
| $α$ (°) | | 90 | 90 | 90 | 90 | 90 | 81.77 |
| $β$ (°) | | 90 | 90 | 90 | 90 | 103.23 | 98.25 |
| $γ$ (°) | | 90 | 120 | 89.16 | 90 | 90 | 89.67 |
| Z | | 2 | 2 | 11 | 2 | 4 | 2 |
| $ρ$ (g/cm$^3$) | | 7.85 | 8.12 | 8.10 | 8.29 | 8.48 | 7.42 |
| $ΔE$ (eV/$Ta_2O_5$) | PBE | 2.326 | 1.966 | 0.309 | 0.243 | 0.117 | 0 |
| | PBEsol | 2.350 | 1.906 | 0.240 | 0.157 | 0.030 | 0 |
| | PW91 | 2.342 | 1.977 | 0.320 | 0.253 | 0.126 | 0 |
| | HSE06 | 2.444 | 2.053 | 0.214 | 0.220 | 0.101 | 0 |
| | PBE0 | 2.506 | 2.042 | 0.273 | 0.207 | 0.018 | 0 |

[a]Expt. [26]; [b]Expt. [24]; [c]Expt. [32].

**TABLE II.** Calculated transition volumes ($V_t^\gamma$, $V_t^B$, $V_t^\lambda$) and the transition pressures of Ta$_2$O$_5$ from the γ phase to B and λ phase.

|  | $V_t^\gamma$ (Å$^3$/f.u.) | $V_t^B$ (Å$^3$/f.u.) | $V_t^\lambda$ (Å$^3$/f.u.) | $P_t$ (GPa) |
|---|---|---|---|---|
| γ → B | 98.08 | 85.46 | --- | 1.78 |
| γ → λ | 97.03 | --- | 86.53 | 4.16 |

**TABLE III**. Geometric parameters describing the radial distribution functions (RDFs) of the λ, B, and γ-$Ta_2O_5$, which are obtained from MD simulations at 300 K (detailed in Supplemental Material). In the case of $g_{TaO}$, the radii $R_{1L}$ ($R_{2L}$), $R_{1U}$ ($R_{2U}$), and $R_{1P}$ ($R_{2P}$) are the lower limit, upper limit and the highest RDF peak positions defining the first (second) coordination shell, respectively. Similar convention applies to the RDF $g_{TaTa}$.

| Radii of coordination shells | $g_{TaO}$ | | | $g_{TaTa}$ | | |
|---|---|---|---|---|---|---|
| | λ | B | γ | λ | B | γ |
| $R_{1L}$ (Å) | 1.72 | 1.68 | 1.72 | 3.08 | 3.10 | 3.10 |
| $R_{1U}$ (Å) | 2.50 | 2.52 | 2.56 | 3.44 | 3.44 | 3.58 |
| $R_{1P}$ (Å) | 1.92 | 1.96 | 1.94 | 3.28 | 3.32 | 3.32 |
| $R_{2L}$ (Å) | 2.50 | 3.10 | 3.38 | 3.44 | 3.44 | 3.64 |
| $R_{2U}$ (Å) | 3.50 | 3.74 | 4.02 | 4.06 | 4.06 | 4.12 |
| $R_{2P}$ (Å) | 3.10 | 3.56 | 3.86 | 3.80 | 3.72 | 3.84 |

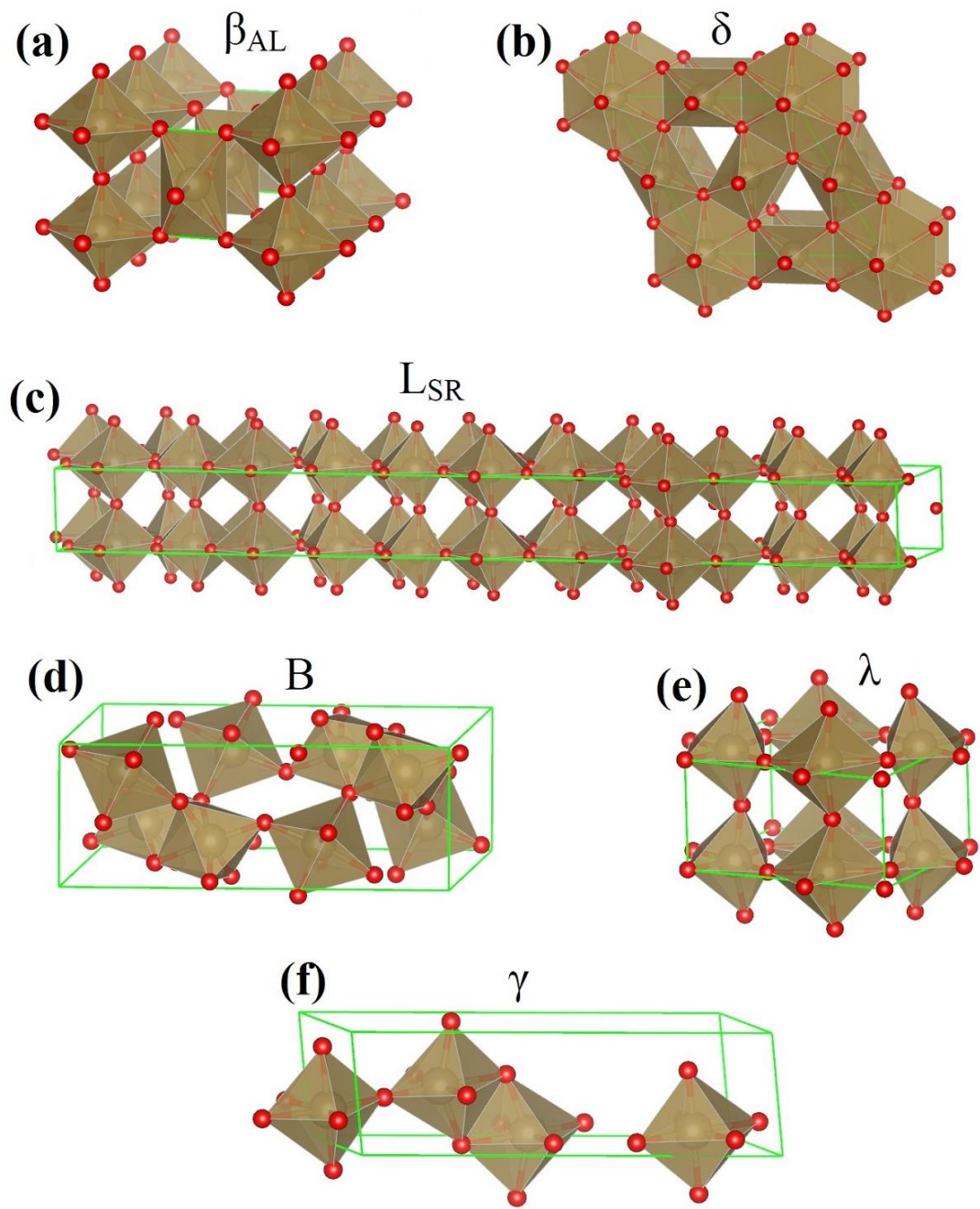

**FIG. 1.** Schematic diagram for the structural models of Ta$_2$O$_5$: (**a**) β$_{AL}$; (**b**) δ; (**c**) L$_{SR}$; (**d**) B; (**e**) λ and (**f**) γ-Ta$_2$O$_5$.

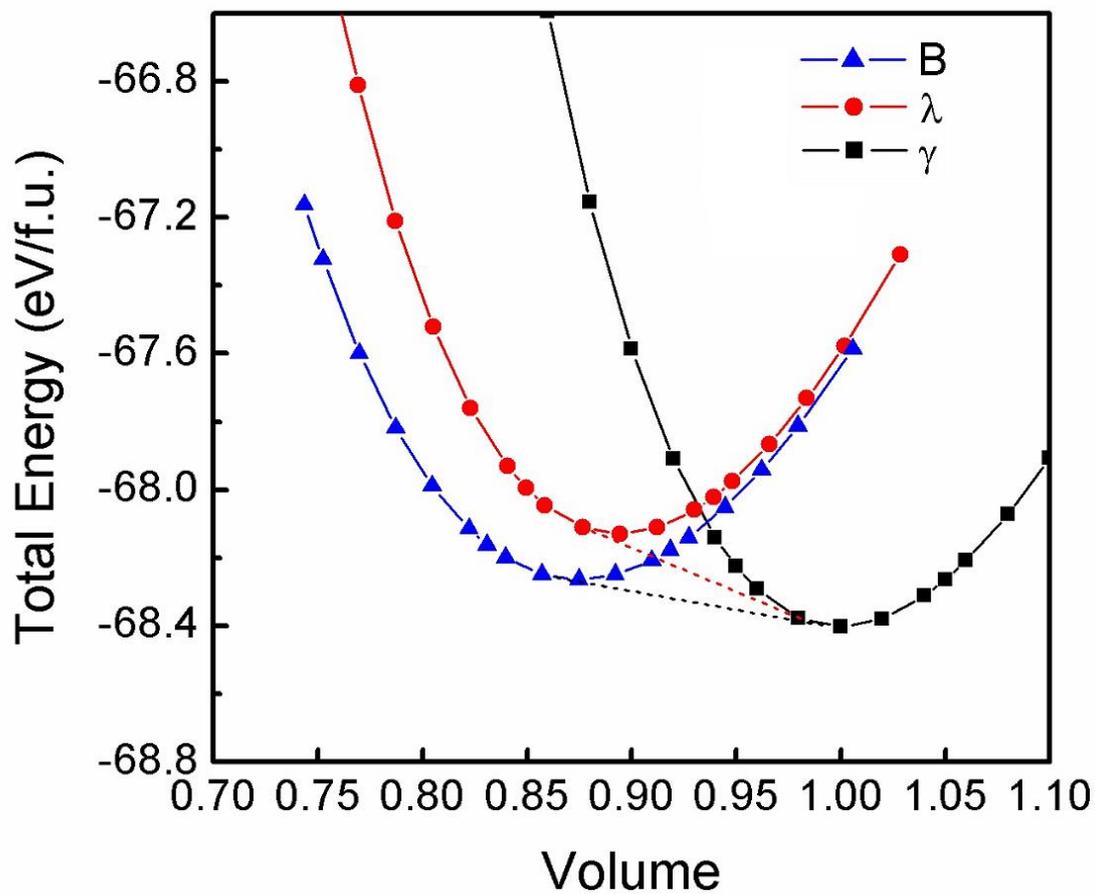

**FIG. 2.** Total energies of the B, λ and γ-$Ta_2O_5$ as a function of volumes normalized to the equilibrium volume of γ-$Ta_2O_5$ at $P = 0$ GPa. The dashed lines are common tangent along which phase transformation between two phases occurs.

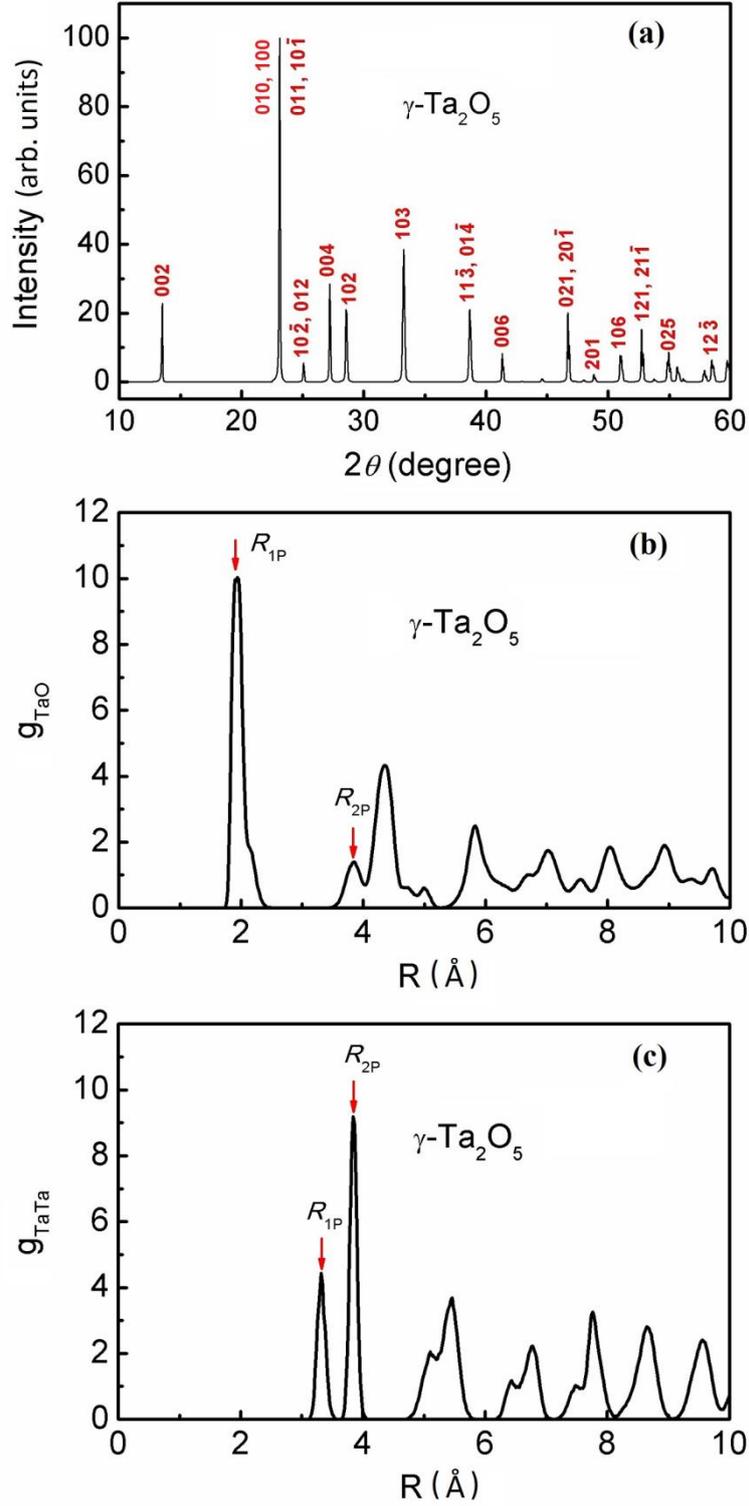

**FIG. 3.** (**a**) Simulated X-ray diffraction (XRD) pattern for γ-Ta$_2$O$_5$, using the Cu *K*α radiation (wave length ~ 1.541 Å). (**b**) – (**c**): The radial distribution functions (RDFs) $g_{TaO}$ (**b**) and $g_{TaTa}$ (**c**) of B-Ta$_2$O$_5$, calculated by averaging over the atomic configurations obtained in MD simulations at 300 K.

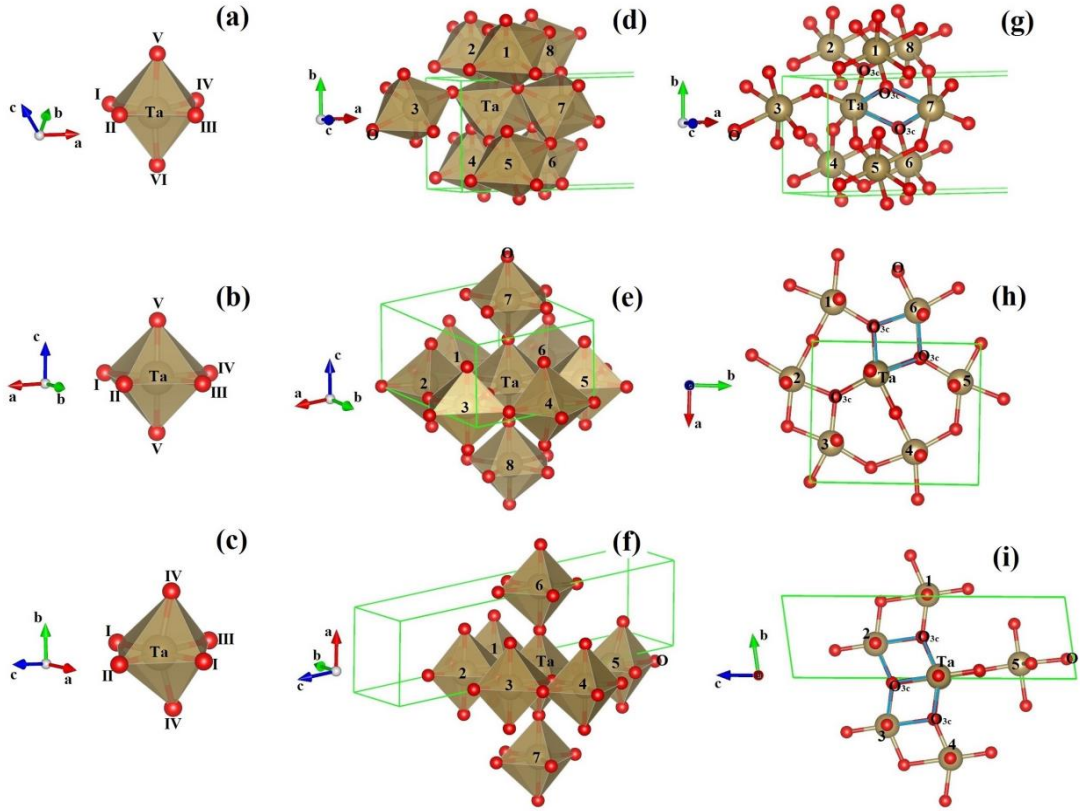

**FIG. 4.** The ball-and-stick representation of TaO$_6$ octahedrons (left panels), their spatial distribution around one central TaO$_6$ (middle panels), and the local bonding structures of O$_{3c}$ (right panels) in B (**a, d, g**), λ (**b, e, h**), and γ-Ta$_2$O$_5$ (**c, f, i**).

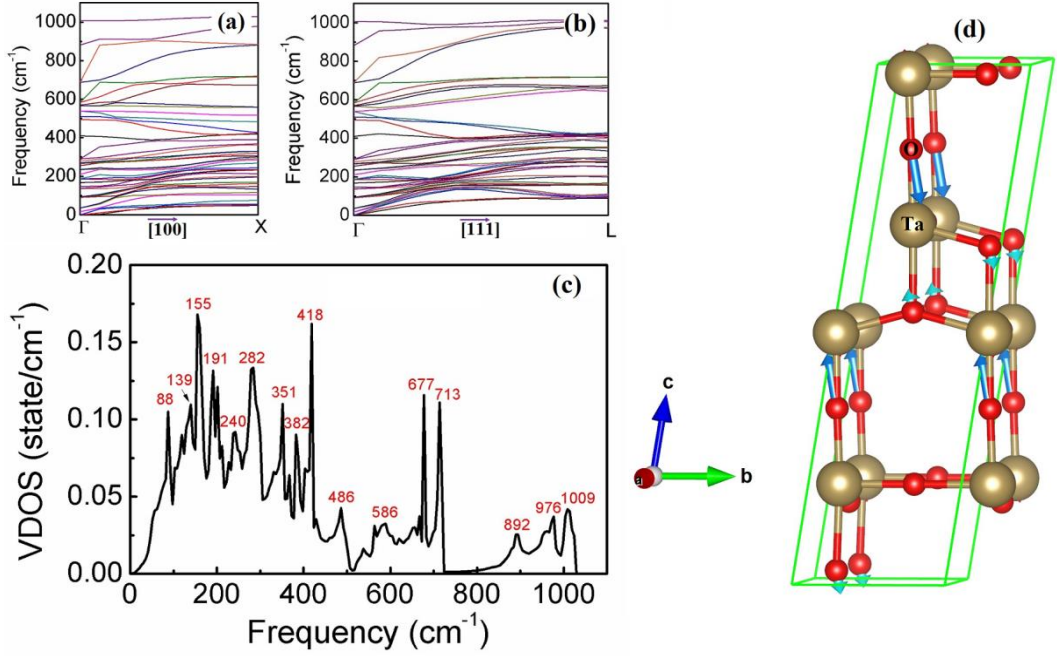

**FIG. 5.** Panels (**a**)-(**c**): Calculated phonon dispersions of γ-Ta$_2$O$_5$ along the ΓX (**a**) and ΓL (**b**) lines; and the vibrational density of states (VDOS, panel **c**). Panel (**d**): Polarization vectors of the vibrational modes of γ-Ta$_2$O$_5$ at Γ-point with the wave numbers $\tilde{v}$ = 1007 cm$^{-1}$. The vibrations of O$_{3c}$ and O$_{2c}$ are represented by sky-blue and deep-blue arrows, respectively.

**Supplemental Material for "Prediction of New Ground State Crystal Structure of $Ta_2O_5$"**

Yong Yang[1*] and Yoshiyuki Kawazoe[2,3]

1. *Key Laboratory of Materials Physics, Institute of Solid State Physics, Chinese Academy of Sciences, Hefei 230031, China.*
2. *New Industry Creation Hatchery Center (NICHe), Tohoku University, 6-6-4 Aoba, Aramaki, Aoba-ku, Sendai, Miyagi 980-8579, Japan.*
3. *Department of Physics and Nanotechnology, SRM University, Kattankulathurm, 603203, TN, India.*

**Contents:**

I. Supplemental Texts
II. Supplemental Tables SI to SIII
III. Supplemental Figs. S1 to S9
IV. References

*E-mail: wateratnanoscale@hotmail.com; yyang@theory.issp.ac.cn

## I. Supplemental Texts

### A. Phonon calculations using Quantum Espresso

Phonon dispersion and vibrational density of states (VDOS) were calculated using Quantum Espresso (QE) package, where the unit cell geometry and atomic structures are fully re-optimized from the results obtained by VASP code [40, 41]. The wave function of valence electrons is expanded using a plane-wave basis set with a kinetic energy cut-off of 75 Ry. The ion-electron interactions are described by the PAW methods [42, 43], and the exchange-correlation energies are described using the Perdew-Zunger functional [66]. For the calculation of the electronic wave function, the Brillouin zone (BZ) is sampled using a 10×10×4 Monkhorst-Pack k-mesh [45]. The Gaussian smearing is employed for doing integration in the Brillouin-zone, with a smearing parameter of 0.05 Ry. A 8×8×2 uniform $q$-point grid is used for the calculation of dynamical matrices, which are then diagonalized to get the eigen-frequency ω ($q$) ($q$: wave vector). The dynamical matrices of denser $q$-meshes can be obtained by using interpolation method and Fourier transform to get the force constants in real space.

After the structural re-optimization using QE package, the obtained unit cell parameters are as follows: $a$ = 3.82 Å, $b$ = 3.82 Å, $c$ = 13.13 Å, $α$ = 81.71°, $β$ = 98.29°, $γ$ = 90.00°. The unit cell is slightly contracted, and the averaged Ta-O bond length decreases from ~ 2.00 Å (by VASP) to ~ 1.97 Å (by QE).

### B. MD simulations using VASP code

To check the thermal stability of γ-$Ta_2O_5$, first-principles molecular dynamics (MD) simulations were performed at 300 K in a system modeled by the (4×4×2) supercell of γ-$Ta_2O_5$, which contains 448 atoms. A canonical ensemble is simulated using the algorithm of Nosé [67-69]. The dynamics of the system are simulated for ~ 2 ps.

MD simulations at 300 K are also performed for B and λ phase of $Ta_2O_5$, in a canonical ensemble for ~ 2 ps, to calculate the averaged radial distribution functions (RDFs) and make a comparison with γ-Ta$_2$O$_5$. A (2×4×3) and (3×3×4) supercell is employed for the simulations of B and λ-Ta$_2$O$_5$, respectively.

For the three simulated systems, the minimum of the lengths of the supercell axis are larger than 15 Å. In all the MD simulations, the time step is 0.5 fs; and only the Γ-point is used for total energy calculations, due to the very large size of simulation systems and consequently much small size of BZ for integration.

## II. Supplemental Tables: TABLE SI to TABLE SIII

**TABLE SI**. Atomic coordinates of γ-Ta$_2$O$_5$ from first-principles calculations, in units of Å.

| Atom | X | Y | Z |
|------|---|---|---|
| Ta1 | -1.20282 | 1.39956 | 9.40698 |
| Ta2 | 2.96610 | 0.54107 | 2.77774 |
| Ta3 | -1.00656 | -0.04931 | 6.40177 |
| Ta4 | -1.26833 | 0.64726 | 13.16289 |
| O1 | 0.88804 | -0.11769 | 6.46959 |
| O2 | -1.03289 | 1.77626 | 7.24990 |
| O3 | 3.05393 | 0.83464 | 0.60284 |
| O4 | 2.89698 | 0.15715 | 4.65301 |
| O5 | -1.06570 | 3.40826 | 9.22737 |
| O6 | 1.04384 | 0.63679 | 2.18186 |
| O7 | 3.01652 | 2.40860 | 3.10127 |
| O8 | 0.78716 | 1.35444 | 9.47984 |
| O9 | -1.21954 | 1.03758 | 11.28834 |
| O10 | -1.19216 | 2.60923 | 13.50713 |

**TABLE SII.** Lattice parameters (unit cell lengths *a*, *b* and *c*) of the six $Ta_2O_5$ phases, calculated using the PBEsol and PW91 exchange-correlation functional (labeled by subscripts). The experimental data are listed for comparison.

|  | $\beta_{AL}$ | $\delta$ | $L_{SR}$ | $\lambda$ | B | $\gamma$ |
|---|---|---|---|---|---|---|
| $a_{PBEsol}$ (Å) | 6.44 | 7.27 | 6.29 | 6.21 | 12.85 | 3.87 |
| $a_{PW91}$ (Å) | 6.51 | 7.33 | 6.34 | 6.25 | 12.93 | 3.88 |
| $a_{expt}$ (Å) | [a]6.217 | -- | [b]6.198 | -- | [c]12.7853 | -- |
| $b_{PBEsol}$ (Å) | 3.68 | 7.27 | 40.56 | 7.34 | 4.87 | 3.87 |
| $b_{PW91}$ (Å) | 3.69 | 7.33 | 40.91 | 7.40 | 4.93 | 3.89 |
| $b_{expt}$ (Å) | [a]3.677 | -- | [b]40.290 | -- | [c]4.8537 | -- |
| $c_{PBEsol}$ (Å) | 7.74 | 3.87 | 3.83 | 3.80 | 5.53 | 13.30 |
| $c_{PW91}$ (Å) | 7.77 | 3.89 | 3.84 | 3.83 | 5.59 | 13.34 |
| $c_{expt}$ (Å) | [a]7.794 | -- | [b]3.888 | -- | [c]5.5276 | -- |

[a]Expt. [26]; [b]Expt. [24]; [c]Expt. [32].

**TABLE SIII.** The bulk modulus and its pressure derivative in Murnaghan's equation of state (EOS), calculated for $\lambda$, B and $\gamma$-$Ta_2O_5$.

|  | $\lambda$ | B | $\gamma$ |
|---|---|---|---|
| $B_0$ (GPa) | 175.0 | 138.6 | 201 |
| $B'$ | 5.3 | 3.8 | 6.43 |

## III. Supplemental Figures: FIG. S1 to FIG. S9

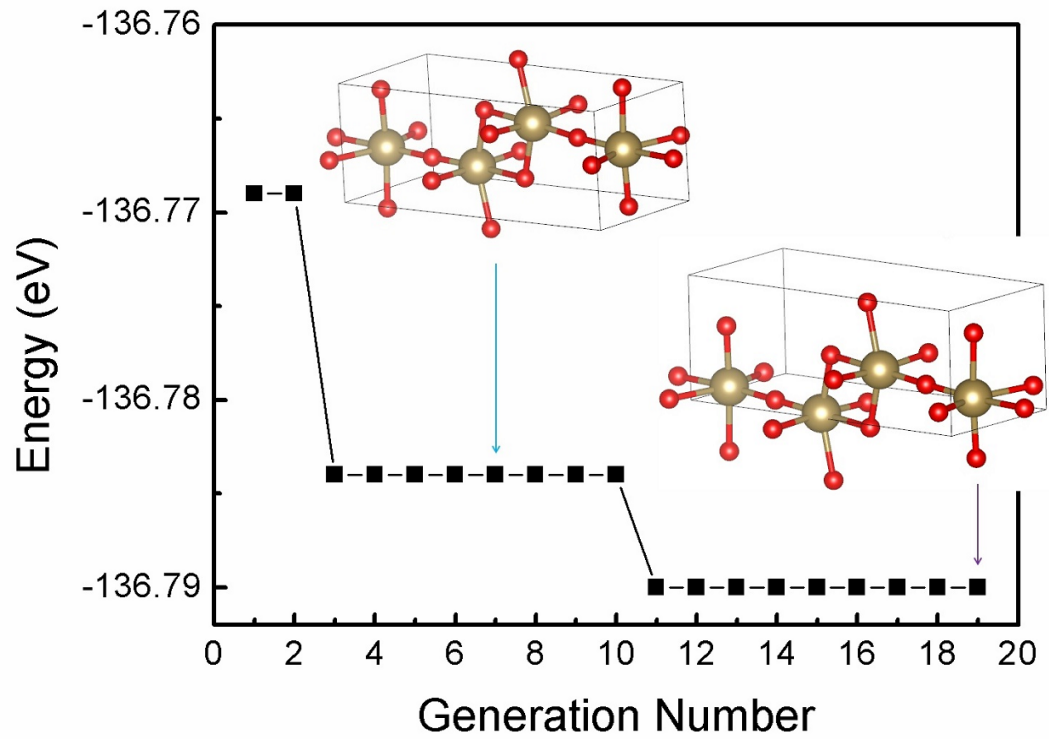

**FIG. S1.** Total energy of the best $Ta_2O_5$ structure as function of generation number predicted by USPEX.

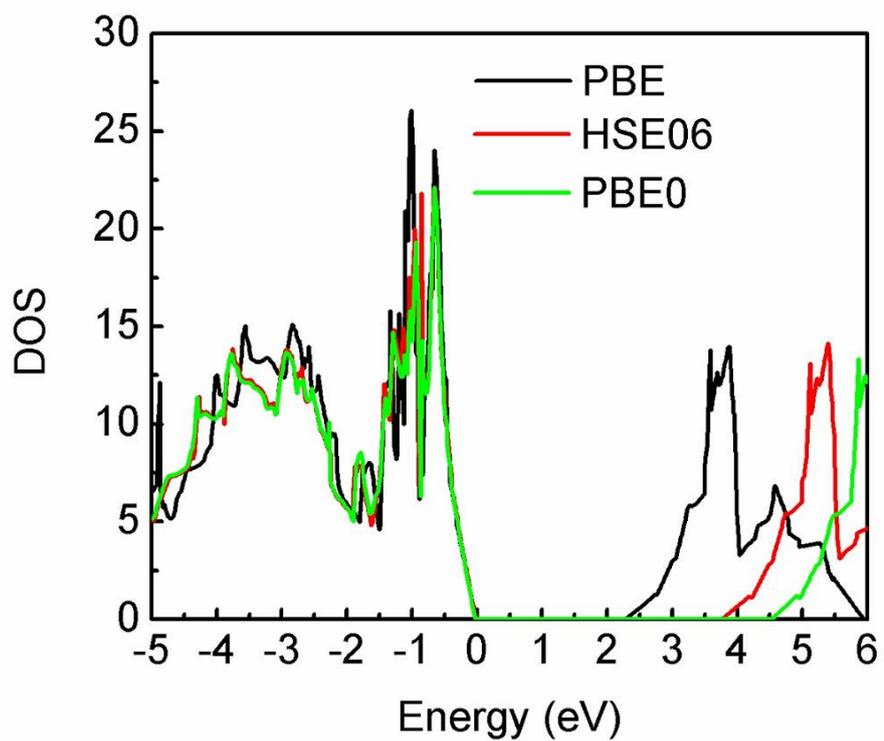

**FIG. S2.** Calculated electron density of states (DOS) of γ-$Ta_2O_5$ using the PBE, HSE06 and PBE0 type functionals. The Fermi level is set at 0.

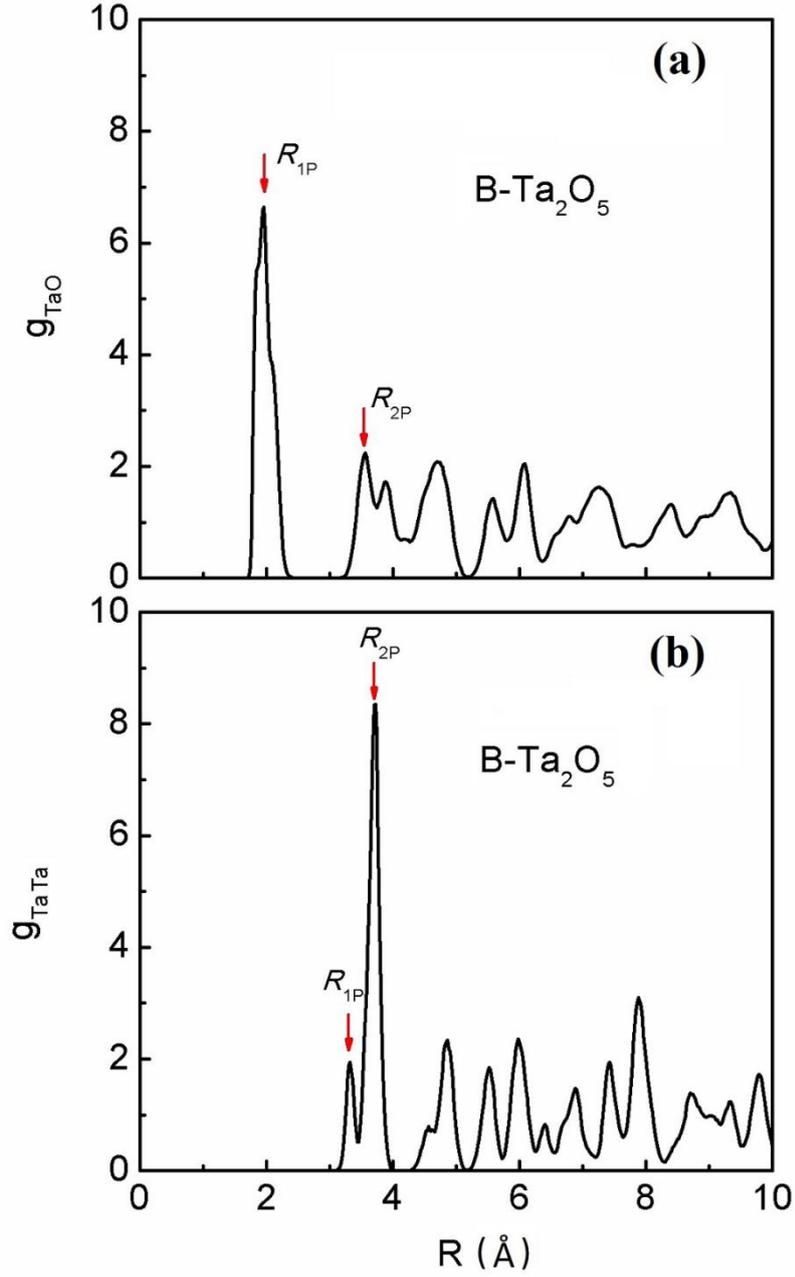

**FIG. S3.** The radial distribution functions (RDFs) $g_{TaO}$ (**a**) and $g_{TaTa}$ (**b**) of B-$Ta_2O_5$, calculated by averaging over the atomic configurations obtained in MD simulations at 300 K.

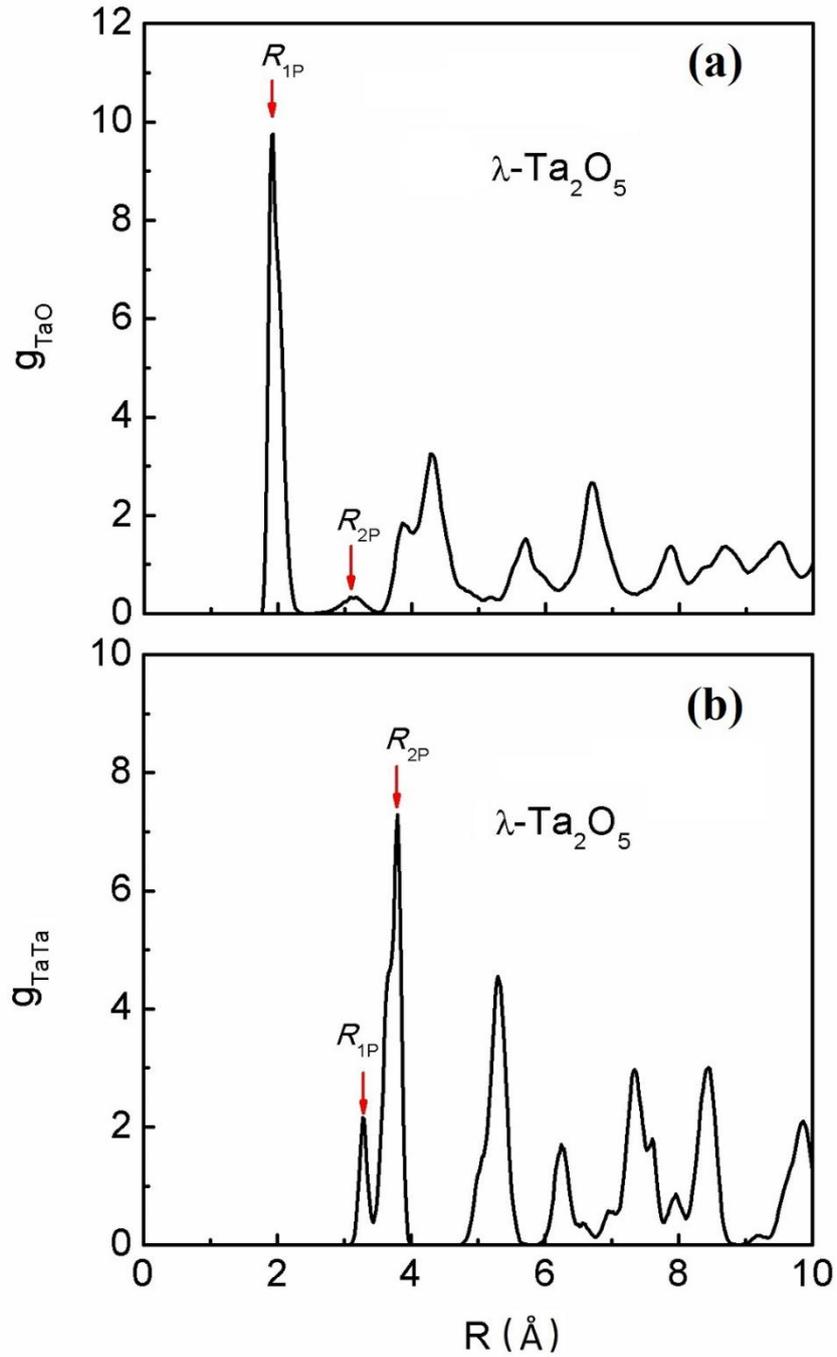

**FIG. S4.** The RDFs $g_{TaO}$ (**a**) and $g_{TaTa}$ (**b**) of λ-$Ta_2O_5$, calculated by averaging over the atomic configurations obtained in MD simulations at 300 K.

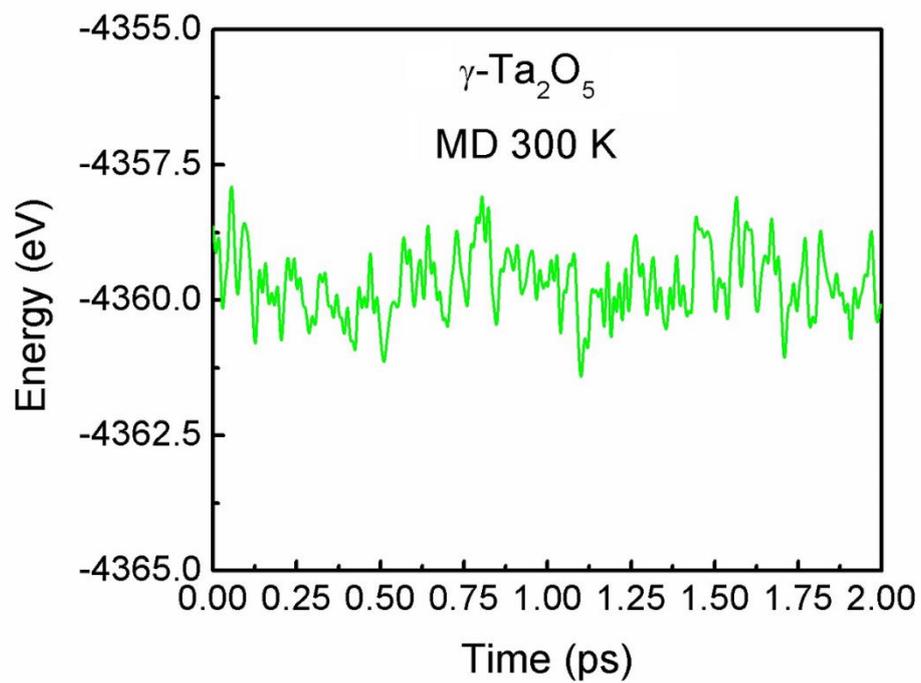

**FIG. S5.** Total energy of γ-Ta$_2$O$_5$ (supercell) as a function of MD simulation time at 300 K.

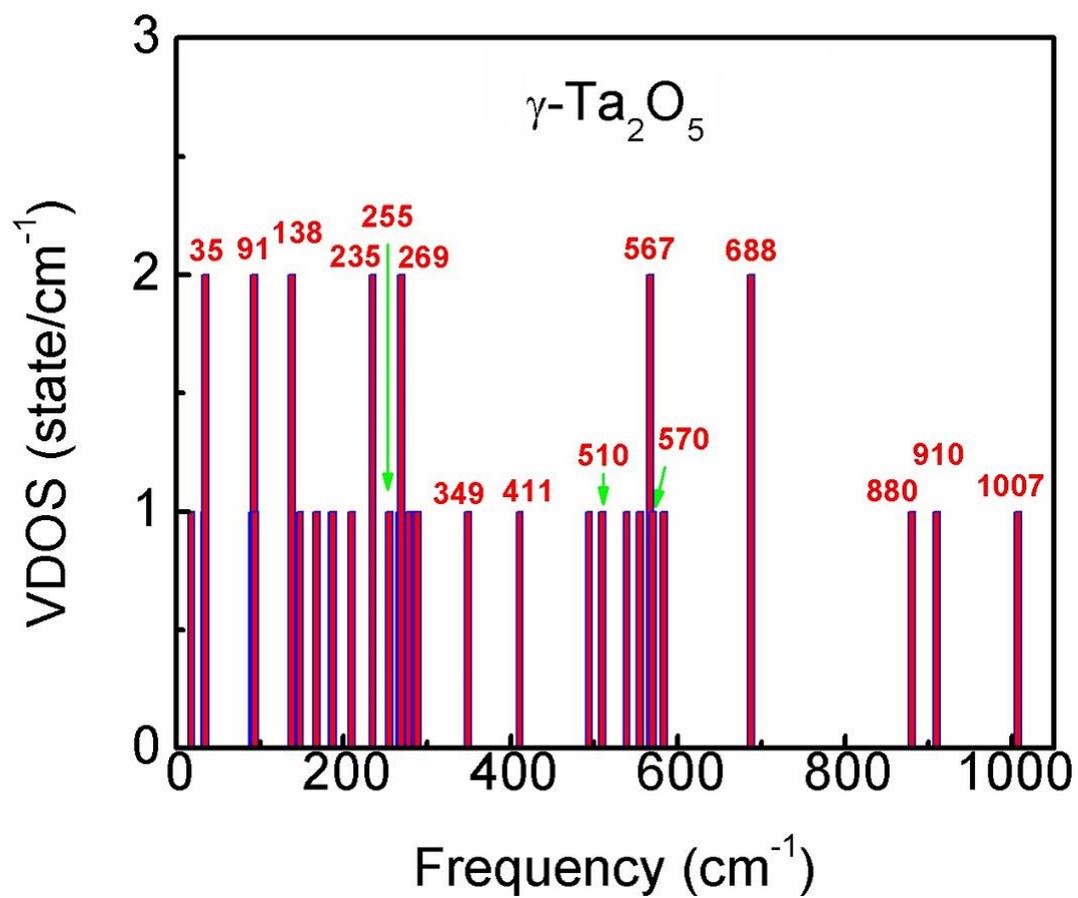

**FIG. S6.** Calculated phonon vibrational density of states (VDOS) of γ-$Ta_2O_5$ at Γ-point.

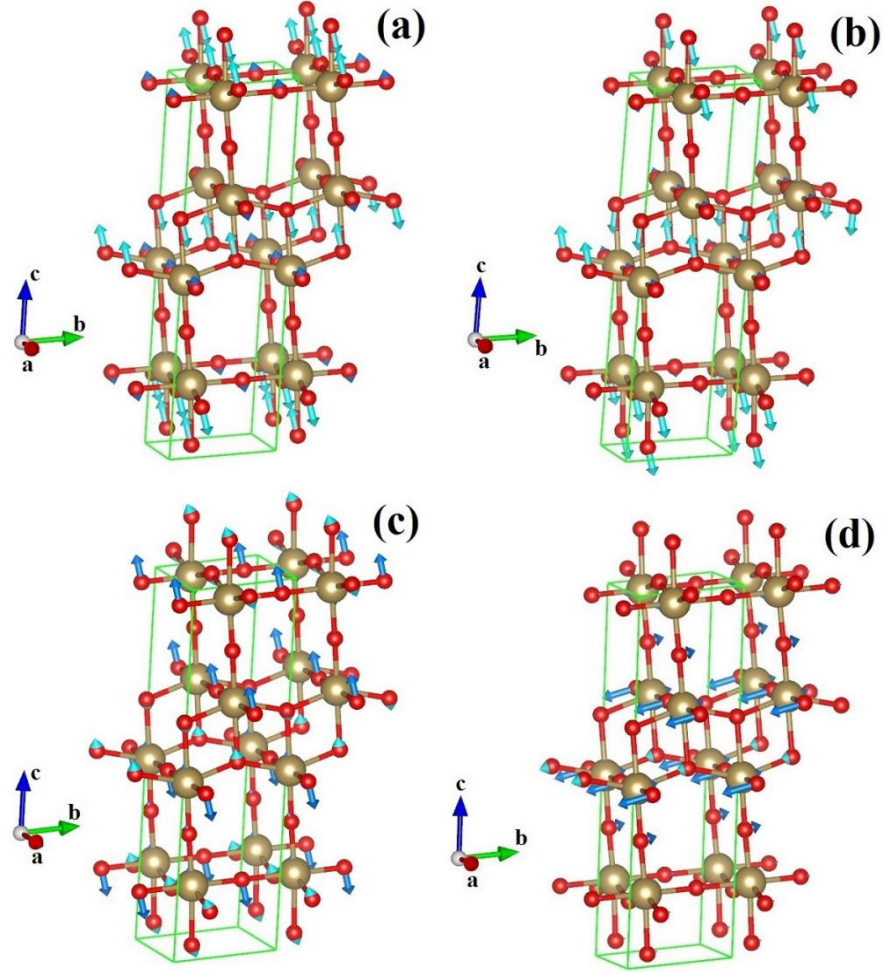

**FIG. S7.** Polarization vectors of the vibrational modes of γ-Ta$_2$O$_5$ at Γ-point with the wave numbers $\tilde{v}$ = 510 cm$^{-1}$ (**a**), 570 cm$^{-1}$ (**b**), 255 cm$^{-1}$ (**c**), 91 cm$^{-1}$ (**d**). The vibrations of O$_{3c}$ and O$_{2c}$ are represented by sky-blue and deep-blue arrows, respectively.

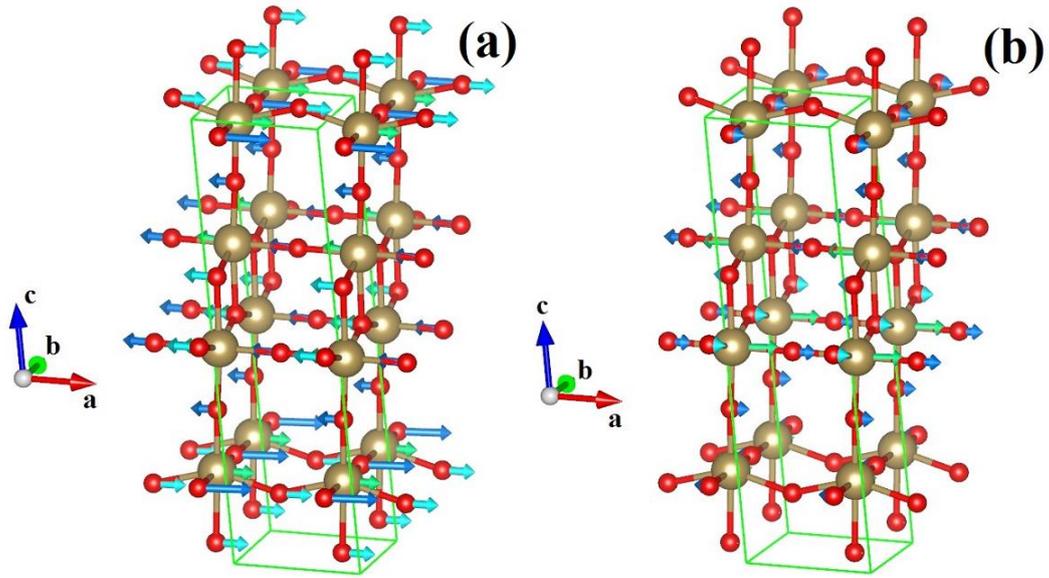

**FIG. S8.** Polarization vectors of the vibrational modes of γ-$Ta_2O_5$ at Γ-point with the wave numbers $\tilde{v} = 34$ cm$^{-1}$ (**a**), 35 cm$^{-1}$ (**b**). The vibrations of $O_{3c}$, $O_{2c}$ and Ta are represented by sky-blue, deep-blue and green arrows, respectively.

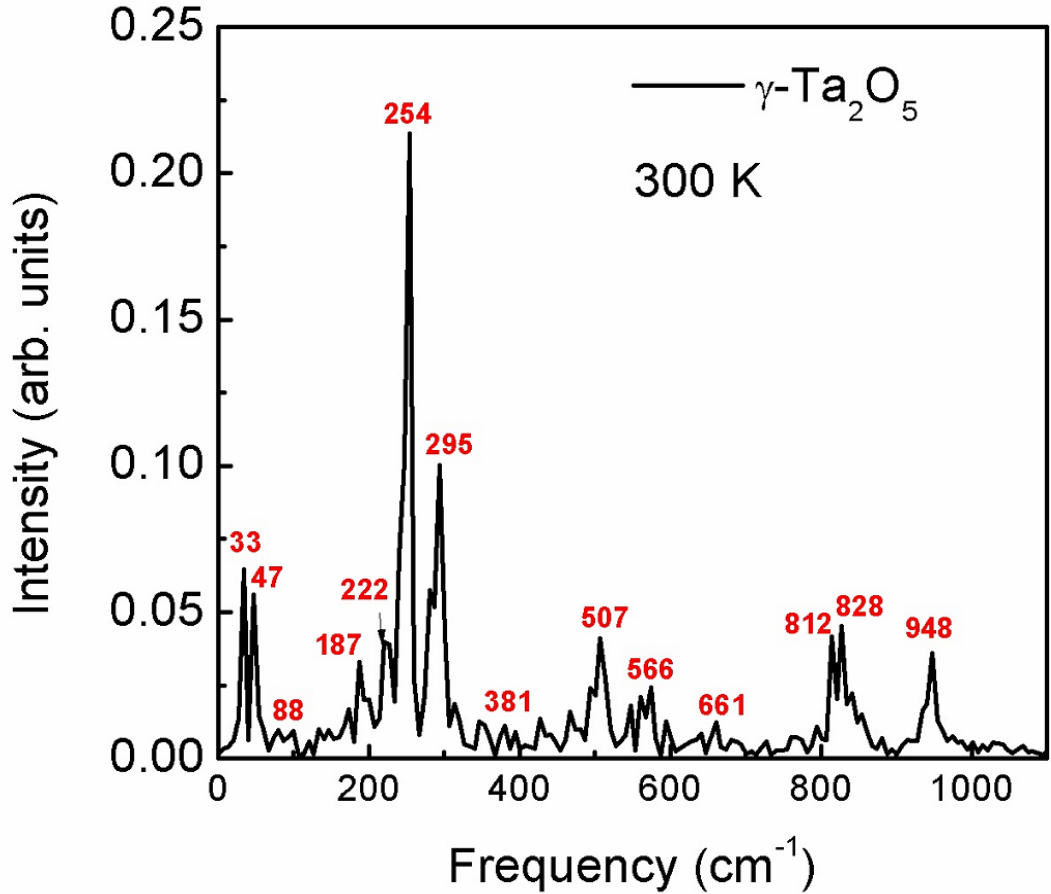

**FIG. S9.** The vibrational spectrum of γ-$Ta_2O_5$, calculated from MD simulations at 300 K using the VASP code.